# Giant energy density nitride dielectrics enabled by a paraelectric-metaparaelectric phase transition

Zhijie Liu[1,2,†], Xingyue Ma[1,2,†], Lan Chen[3], Xiaohong Yan[3], Jun-Ming Liu[1,2], Chun-Gang Duan[4,5*], Jorge Íñiguez-González[6,7], Di Wu[1,2], Yurong Yang[1,2*]

[1]*Laboratory of Solid State Microstructures, Nanjing University, Nanjing 210093, China*

[2]*Jiangsu Key Laboratory of Artificial Functional Materials, Nanjing University, Nanjing 210093, China*

[3]*School of Science, Nanjing University of Posts and Telecommunications, Nanjing, 210023, China*

[4]*Key Laboratory of Polar Materials and Devices (MOE) and State Key Laboratory of Precision Spectroscopy, East China Normal University, 500 Dongchuan Rd., Shanghai 200241, China*

[5]*Collaborative Innovation Center of Extreme Optics, Shanxi University, Taiyuan 237016, China*

[6]*Department of Materials Research and Technology, Luxembourg Institute of Science and Technology, 5 Avenue des Hauts-Fourneaux, L-4362 Esch/Alzette, Luxembourg*

[7]*Department of Physics and Materials Science, University of Luxembourg, 41 Rue du Brill, L-4422 Belvaux, Luxembourg*

[†] These authors contributed equally to this work.

*Corresponding author. Email: cgduan@clpm.ecnu.edu.cn (C-G.D.); yangyr@nju.edu.cn (Y.Y.)

**Abstract**:

Electrostatic dielectric capacitors are foundational to advance the electronics and electric power devices due to their ultrafast charging/discharging capability and high-power density. However, the low energy density limits the potential for next generation devices in terms of miniaturization and integration. We propose a strategy that relies on inducing a field-driven phase transition that we denote paraelectric-metaparaelectric, which yields an ultrahigh energy density in III-nitrides. III-nitride compounds (Al, Sc, B)N with certain cation concentrations possess a nonpolar hexagonal ground phase which could transform into a polar wurtzite phase under a very large electric field, which is denoted as metaparaelectric with nearly null hysteresis *P-E* loop. This paraelectric-metaparaelectric transition leads to a polarization saturation at large electric field. The corresponding *P-E* loop displays a giant energy density of 308 J/cm$^3$

---

with high efficiency nearly 100%. The proposed paraelectric-metaparaelectric phase transition strategy in nitrides opens an avenue to design of next generation high performance dielectrics.

## Introduction

Electrostatic energy storage based on dielectric capacitors have broad application in electric power systems and electronic devices, due to their ultrafast charge and discharge and ultrahigh power density[1-3]. However, comparing to the electrochemical energy-storage technologies, the energy density of the dielectric capacitors is generally low, which limits the application and further development in terms of device miniaturization and integration. Thus, extensive efforts have been focusing on the development of reliable and efficient dielectric materials with high energy density.

The energy density $W$ of dielectric capacitors can be defined as $W = \int_{P_r}^{P_m} EdP$, where $P_m$ and $P_r$ represent the maximum polarization after charge and remnant polarization after discharge[4], respectively (see Fig. 1). The energy efficiency $\eta$ is defined as $\eta = W/(W+W_{loss})$, where $W_{loss}$ is the loss in the discharging process due to hysteresis. To obtain high energy density $W$ and energy efficiency $\eta$, the large polarization variation $\Delta P = (P_m - P_r)$, large breakdown electric field ($E_b$), and small hysteresis are desired[5]. Linear dielectrics usually exhibit low maximum polarization ($P_m$), while ferroelectrics (FEs) exhibit large remnant polarization ($P_r$) (see Fig. 1a); therefore, both linear dielectrics and FEs usually possess low energy density ($W$). Relaxor-ferroelectrics (RFEs) (see Fig. 1b) are proposed for high energy density by introducing nanodomains in ferroelectrics via composition and defect engineering[6-11]. The nanodomains induce lower remnant polarization ($P_r$) and reduce the polarization switching barrier, leading to and smaller hysteresis ($W_{loss}$), which achieve an energy density of 112 J/cm$^3$ in ferroelectric film[6]. Superparaelectric (SPE) RFEs with minimal hysteresis and very high efficiency and large energy density of 152 J/cm$^3$ were realized by scaling down the nanodomains to several unit cells[12]. However, one restriction of RFEs is that they achieve a polarization saturation limit ($P_m$) at a low electric field, which leads to low energy density ($W$). Relaxor antiferroelectrics (AFEs) are an alternate to improve the saturation electric field, because of their characteristic of nonpolar-polar phase transition at a large electric field[13,14] (see Fig. 1c). Unfortunately, the number of AFE materials are limited and most of them exhibit small breakdown strength[15-18]. Though the energy density of 200 J/cm$^3$ was achieved by carefully preparing the samples in RFEs or Relaxor AFEs[8], it is still significantly lower than that of the electrochemical battery, highlighting the need of a strategy and materials to supplement the previous methods and lead to high performance of electrostatic capacitors with high energy density.

Here, by first principles calculations, we introduce a strategy for controlling the *P-E* loop which maintains a polarization saturation limit at large electric field and a minimal hysteresis, via what we call a paraelectric-metaparaelectric (PE-MPE) phase transition (see Fig. 1d) in nitride dielectrics, whose features are described below. Based on the fact that wurtzite nitrides of Sc-doped or B-doped AlN are ferroelectric materials with large polarization (>100 μC/cm$^2$) and strong breakdown strength (>6 MV/cm)[19-25], we extensively study the III-nitride compound (Al, Sc, B)N. We identify compounds present a PE ground state that can transform into ferroelectric or MPE phase under electric field. The PE-MPE phase transition results in a giant energy density up to 308 J/cm$^3$ and an ultrahigh efficiency of 98%. The value of energy density is much larger than the value 202 J/cm$^3$ from previous reference[8]. We therefore believe that the proposed strategy and nitride compound have the potential of platform for dielectric capacitor that possesses high energy density and efficiency.

## Results

**Structure stability and phase diagram**

Binary III-nitrides have five possible structural phases. As shown in Fig. 2, these structures are wurtzite (WZ), hexagonal (HE), rock salt (RS), zinc blende (ZB)[26], and β-BeO structure (BB)[27]. WZ, RS, and ZB are general stable phases; the ground state for AlN, BN, and ScN is WZ, ZB, and RS, respectively. WZ is the only polar phase which possesses large polarization (134 μC/cm$^2$ for AlN, 213 μC/cm$^2$ for BN). The HE phase (BB phase, respectively) is the metastable transient structure during the polarization switching for ferroelectric Al$_{1-x}$Sc$_x$N (Al$_{1-x}$B$_x$N, respectively)[23,24]. For bulk

AlN, the HE and BB phases have higher energy than the WZ phase by 183.5 meV per formula unit (f.u.) and 76.6 meV/f.u., respectively (see Table SI in supplementary materials). The previous research shows that the minimum energy barrier of switching polarization for WZ phase of AlN is 192 meV/f.u[24]. This energy barrier is too high to switch polarization in bulk WZ AlN by electric field; for this reason, doping AlN with Sc and B is used to lower the energy barrier for ferroelectric polarization switching[23,24]. For instance, the energy barrier can be reduced to 60 meV/f.u at 32% Sc concentration. Two factors are considered to be important to influence polarization switching in AlN-based materials: 1) doping concentration of Sc, as the energy barrier and electric field required for polarization switching is decreased for increasing Sc concentration[19,24]; 2) strain imposed by the substrate. The calculated and experimental result show that in-plane tensile strain can reduce the coercive field and energy barrier[38,39].

The paraelectric HE and BB phases, which were observed as the intermediate nonpolar phases during the ferroelectric switching of WZ phase[23,24], can transform into the WZ polar state with a high polarization under electric field[28]. This nonpolar-polar phase transition is similar to the AFE-FE phase transition for AFEs under electric field, enabling storage of a high energy density because of the *P-E* characteristic curve. In this *P-E* curve the polarization reaches its saturation limit at large electric field; this contrasts with the case of RFEs, where the *P-E* curve reaches saturation at relatively small fields which results in a reduced energy density. The HE and BB phases are thus good candidates to obtain high energy density dielectrics, provided the materials we can find that are the ground state.

Making the HE or BB phases the ground state thus becomes the key factor to design high energy density dielectric capacitors based on III-nitrides. Doping the WZ phase of AlN is a valid strategy to tune the relative stability of the different structural polymorphs. Sc-doped AlN [(Al, Sc)N] can present WZ and RS ground states under different Sc concentrations (see Fig. S1a)[24,29]. B-doped AlN [(Al, B)N] can present WZ and ZB ground states under different B concentration[23,24] (see Fig. S1b). Single doping of Sc or B cannot yield a HE or BB ground state, though. We then manipulate the structural stability and phase diagram of III-nitride compound (Al, Sc, B)N by varying the concentration of three cations.

The stability of nitride phases can be quantified by their formation enthalpy, $\Delta H_f = E_{tot} - xE_{ScN} - yE_{BN} - (1-x-y)E_{AlN}$, where $E_{tot}$ is the total energy per formula unit of $(Al_{1-x-y}Sc_xB_y)N$, $x$ ($y$, respectively) is the doping concentration of ScN (BN, respectively), $E_{ScN}$ ($E_{BN}$) is the ground state energy of ScN (BN, respectively), and $E_{AlN}$ is the ground state energy of AlN. Note that B is much smaller than Al and Sc; hence, the structure of ternary nitrides (Al, Sc, B)N is distorted near the B cations compared to simple single-cation nitrides (see Fig. 2 and Fig. 3d), and we denote the resulting structural phases of (Al, Sc, B)N with a prime (WZ′, HE′, BB′, RS′, ZB′). In order to illustrate trends of phase stability, the change of formation enthalpy for a B concentration of 22% is selected as an example. Figure 3a shows the formation enthalpy as a function of Sc concentration $x$ for $(Al_{0.78-x}Sc_xB_{0.22})N$. The results show the ground state of the III-nitride compound is ZB for $x$=0, same as the ground state of BN (see Table S1). Increasing $x$ from 3% to 35%, the ground state is WZ′ (see Fig. 3d). And further increasing $x$ from 38% to 63%, the ground state changes to nonpolar HE′ (see Fig. 3d), which can transform into polar WZ′ under an electric field and could thus yield a high energy density. The ground state becomes RS′ (see Fig. 3d) when the Sc concentration $x$ increases from 66% to 78%, same as the ground state of ScN (see Table S1).

The HE′ phase found by increasing the Sc concentration in systems with a B concentration of 22% points at a promising possibility for realizing large energy storage density. To further investigate the concentration range where HE′ is the ground state, the whole concentration range of Al, Sc and B has been studied. We thus obtain a ternary phase diagram of (Al,Sc,B)N, as shown in Fig. 3b. The phase diagram can be approximately divided into four ground state regions dependent on the Sc concentration

*x*. They are sequentially ZB′, WZ′, HE′, and RS′ when increasing the Sc concentration *x* from 0 to 100% (also see Fig. 3b), roughly independent from the B concentration (except for very small B content). For small Sc concentration of $0 \leq x \leq 9\%$, ZB′ is the ground state when the B concentration is very large (blue region in Fig. 3b) and WZ′ is the ground state when the B concentration is small (orange region in Fig. 3b). For Sc concentrations $13\% \leq x \leq 38\%$, the ground state is mostly the WZ′ phase. Further increasing the Sc concentration to $38\% < x \leq 66\%$, HE′ becomes the ground state for most B concentrations (the green region in Fig. 3b). For $x > 66\%$, RS′ is the ground state (the purple region in Fig. 3b).

The WZ′ and HE′ phases have similar lattice symmetry and can be transformed into each other by an external field or by changing the Sc concentration *x*. The relative stability between the ferroelectric (polar) WZ′ phase and the paraelectric (nonpolar) HE′ phase is reversed the as *x* increases. As shown in Fig. 3c, the ferroelectric WZ′ phase has much lower energy than paraelectric HE′ phase for small Sc concentration *x*. Then, the ferroelectric WZ′ phase gradually increases its energy and transforms into the HE′ phase when *x* grows to about 38% (the orange and green color boundary in Fig. 3a). Further, then HE′ phase may transform to WZ′ phase when increasing the polarization large enough (green line in Fig. 3c). This interesting energy hierarchy between ferroelectric and paraelectric phases suggests the possibility of having phase transitions that could be leveraged to improve the stored energy density.

### *P-E* response and energy storage capability

Based on the unique *P-E* loop of PE-MPE phase transition shown in Fig. 1d, the paraelectric HE (or HE′) phase may allow us to store a large energy density. We thus study the electric polarization under electric field and energy storage performance for HE′ phase by first-principles calculations. (The calculations are done in the limit of 0 K, but they offer relevant information about the relevant potential energy surface.) Figures 4a and 4b show the *P-E* loops for $(Al_{1-x-y}Sc_xB_y)N$ with $x = 50\%$ and 59%, respectively. For each of the Sc concentrations five *P-E* loops are calculated, for B concentration *y* from 13%-25%. In Fig. 4a, for Sc concentration of 59%, the variation of polarization with respect to initial structure *P* is up to about 88 $\mu C/cm^2$ for a B concentration of 13%. With the increase in B concentration from 13% to 25%, the value of *P* gradually decreases to about 57 $\mu C/cm^2$, implying that larger *y* in $(Al_{1-x-y}Sc_xB_y)N$ leads to smaller *P*. For the Sc concentration of 50% (see Fig. 4b), the largest *P* value is 86 $\mu C/cm^2$ at the B concentration of 13%, and the smallest *P* is 54 $\mu C/cm^2$ for the B concentration of 25%. Compared to the large Sc concentration $x = 59\%$, reducing *x* to 50% yields a smaller *P*. We thus observe that *P* increases for increasing Sc content and decreasing B content; this result is essentially driven by the relatively large Born effective charge of 2.9*e* for Sc and the relatively small Born effective charge of 2.3*e* for B, compared to 2.7*e* for Al for the wurtzite structure. The *P-E* loops of HE′ phase of $(Al_{1-x-y}Sc_xB_y)N$ have distinct characteristics compared to traditional RFEs *P-E* loops, as they i) achieve the saturation of *P* limit at large electric field (in our calculations, some of them reach the saturation at electric field larger than 6 MV/cm which is not shown in Fig. 4), ii) display very small hysteresis with small energy loss, and iii) present a very large *P*. These characteristics would benefit the energy density of dielectric materials.

Figure 4c displays the energy density and the corresponding efficiency for $(Al_{1-x-y}Sc_xB_y)N$ $(x = 47\% \sim 59\%, y = 13\% \sim 25\%)$. The energy density increases with an increasing Sc concentration and decreases when the B content grows. The energy density varies between 160 $J/cm^3$ and 310 $J/cm^3$ for different *x* and *y* with a maximum applied field of 6 MV/cm. And the efficiencies are all above 90%, some being almost 100%, much higher than most of the reported dielectrics. Strikingly, we find that for B concentrations of 13% and 16%, the III-nitride compounds have a giant energy density exceeding 220 $J/cm^3$, which had never been achieved before. For $(Al_{0.28}Sc_{0.59}B_{0.13})N$, the predicted energy density is as large as 308 $J/cm^3$ with an efficiency of 98%, which is even close to the density of an electrochemical battery. Figure 4d further shows the energy density of $(Al_{1-x-y}Sc_xB_y)N$ as a function of *x* and *y*. One can see that the HE′ phase could display large energy densities, the (Al$_{0.28}$Sc$_{0.59}$B$_{0.13}$)N composition showing the largest result. Please note that (Al, Sc)N and (Al, B)N possess a breakdown strength larger than 6.0 MV/cm (see Fig. S2b)[21,22], suggesting that the breakdown electric field of (Al, Sc, B)N could be above 6.0 MV/cm, which could enable energy densities beyond 308 J/cm$^3$ to be stored in (Al, Sc, B)N.

The giant energy density of the HE′ nitride (Al, Sc, B)N can be understood by inspecting the transition between the paraelectric phase and the state that we denote metaparaelectric (MPE). In (Al, Sc)N, the nonpolar HE phase is a high-lying meta-stable polymorph (see Fig. S1 and Ref [24]). The HE′ phase is favored as Al is gradually replaced by B, to the extent that it is predicted to become the ground phase when the B concentration is large enough (see green region in Fig. 3b). For example, HE′ is a meta-stable phase for (Al$_{0.41}$Sc$_{0.59}$)N, while it becomes the ground phase for (Al$_{0.28}$Sc$_{0.59}$B$_{0.13}$)N (see Fig. S1 and Fig. 3). In order to illustrate the defining features of the PE-MPE transition, Figure 5a shows illustrative examples for the *P-E* loops of HE phase in (Al$_{0.41}$Sc$_{0.59}$)N and HE′ in (Al$_{0.28}$Sc$_{0.59}$B$_{0.13}$)N. *P-E* loop of (Al, Sc)N clearly displays a PE-FE phase transition with a distinct hysteresis and coercive electric field of about 2 MV/cm. By contrast, in the *P-E* loop corresponding to the HE′ phase of (Al, Sc, B)N, the polarization increases linearly for small electric field, to then grow more quickly – but also in a quasi-linear fashion – starting from 4 MV/cm. Above 4 MV/cm, (Al, Sc, B)N presents the WZ′ phase, which is characterized by a relatively large distortion compared to the WZ polymorphs of (Al, Sc)N (see Fig. 5b and c). The difference between the two *P-E* loops in Fig. 5a can be traced back to the Born effective charges of Sc and B. As Sc in the HE phase has a larger Born effective charge (3.9 e) than B in the HE′ phase (1.5 e), the HE phase of (Al, Sc)N (blue curve in Fig. 5a) has a much larger *P* than (Al, Sc, B)N, and could transform into the WZ structure at a smaller coercive field. The small Born effective charge of B in the (Al, Sc, B)N compound leads to a smaller dielectric constant; in other words, *P* of (Al, Sc, B)N (red curve in Fig. 5a) under electric field increases slowly, which ultimately results in a larger coercive field being needed to transform into the WZ′ phase. Figure 5b and c show the corresponding atomic structures for the *P-E* loops. At zero field, (Al,Sc)N is a HE phase with cations and N ions almost sitting in the same (001) planes (see left panel of Fig. 5b). Under an electric field of 3 MV/cm, we obtain the WZ phase where the cations and N ions no longer sit in the same (001) planes (see middle panel of Fig. 5b). Further increasing the electric field to 4 MV/cm, the structure of the WZ phase changes only slightly (right panel of Fig. 5b). By contrast, in the case of (Al, Sc, B)N (Fig. 5c), the Al, Sc, B and N atoms sit close to the same (001) plane at zero field, but with some distortions (mainly around B atoms due to its small size). As the field increases to 3 MV/m, the cations move slightly along the *c* axis, the movements being much smaller than those in (Al, Sc)N (middle panels in Fig. 5b and c), indicating that the material is basically still in the HE′ phase. As the electric field further increases to 6 MV/cm, the sample transforms into the WZ′ phase with a large polarization (right panel of Fig. 5c). This large distortions around the B atoms effectively yield a relatively small energy barrier between the HE′ and WZ′ phases of for (Al, Sc, B)N, which results in a similarly tiny hysteresis (see Fig. 5a). Indeed, in the regime between 4 MV/cm and 6 MV/cm, the *P-E* curve is approximately linear, and the hysteresis is practically null, suggesting that the system still behaves as if it were in a PE state. This paraelectric-like WZ′ state is strongly polarized, yet its polarization does not show signs of saturation. We propose the term "metaparaelectric" to denote this quasi-linear and reversible regime of *P-E* curve at high applied fields, emphasizing that its occurrence is ideal for the optimization of the stored energy density.

In general B doping of (Al, Sc)N nitride is predicted to have the following effects: i) it leads the HE′ phase to be the ground state of the system, yielding a small $P_r$; ii) it increases the critical electric field at which the PE-FE or PE-MPE transition occurs, which in turn pushes up the saturation field and the energy density that can be stored; iii) it shrinks the hysteresis, reducing the energy loss. All these three effects benefit the dielectric energy density, making the energy storage capability of (Al, Sc, B)N beyond 300 J/cm$^3$, much larger than those of reported oxide dielectrics.

## Discussion

To further understand the mechanism of the high energy storage performance, the Landau-type model is employed, where the free energy is written as

$$F = \frac{1}{2}aP^2 + \frac{1}{4}bP^4 + \frac{1}{6}cP^6 + \frac{1}{8}dP^8 - EP, \tag{1}$$

where $a$, $b$, $c$ and $d$ are coefficients of quadratic, quartic, sextic and octic terms, respectively. Under equilibrium conditions, the free energy satisfies $\frac{\partial G}{\partial P} = 0$, which leads to

$$E = aP + bP^3 + cP^5 + dP^7. \tag{2}$$

The energy density then can be calculated as

$$\begin{aligned} W &= \int_{P_r}^{P_m} E dP = \int_{P_r}^{P_m} (aP + bP^3 + cP^5 + dP^7) dP \\ &= \frac{1}{2}a(P_m^2 - P_r^2) + \frac{1}{4}b(P_m^4 - P_r^4) + \frac{1}{6}c(P_m^6 - P_r^6) + \frac{1}{8}d(P_m^8 - P_r^8), \end{aligned} \tag{3}$$

where $P_r = 0$ in this study, as the polarization under zero electric field and the polarization when removing electric field in the *P-E* curve are the same.

As shown in Fig. 5a, the model Eq. (2) fits *P-E* loops of the first principles calculations very well, and Eq. (3) could describe the energy density accurately. To compare the energy storage performance of considered III-nitride with the RFEs oxides, Figure 5a also shows the *P-E* loop of RFE oxide from Ref. [8] which exhibit high energy density with rescaled $E_b$ to 6 MV/cm and $P$ to 80 μC/cm². Table I shows the parameters of *a, b, c, d*, and *W* [for Eq. (2) and Eq. (3)] for $(Al_{0.41}Sc_{0.59})N$, $(Al_{0.28}Sc_{0.59}B_{0.13})N$ and RFE oxide. One thus can confirm that *a* for PE-MPE is larger than the others, implying that in the corresponding electric filed (*E*) vs polarization (*P*) curves (see Fig. S4a) *E* for PE-MPE increases faster than that PE-FE and RFE. *b* is negative for PE-FE, which leads *E* to increase very slowly and even decrease at $P = 40$ μC/cm². By contrast, *b* is positive and large for RFE which leads *E* increasing very fast at about $P = 40$ μC/cm². *b* is very small for PE-MPE which does not affect the slope of the *E-P* curve (see Fig. S4a). Then, *c* is negative for PE-MPE, which leads to *E* increase very slowly at large *P*.

Comparing the three curves in Fig. 5a (or Fig. S4), we have that the large $a$, small $b$, and negative $c$ are key to the unusual non-linear behavior of the PE-MPE transition and to thus obtain a giant energy density.

From Eq. (1) and the fitting parameters of Eq. (2), we also calculated the Landau free energy (F) as a function polarization ($P$) for $(Al_{0.41}Sc_{0.59})N$, $(Al_{0.28}Sc_{0.59}B_{0.13})N$ and RFE oxide under zero field, as shown in Fig. S4b. The $F$-$P$ curve for PE-FE is similar to the green curve in Fig. 3c, which is consistent with a PE-FE transition and hysteretic behavior. The $F$-$P$ curve for PE-MPE is similar to the blue curve in Fig. 3c, corresponding to a situation where the PE-FE transition does not fully occur (in the considered range of electric fields) and the system is essentially hysteresis free.

In summary, we extensively study the structural stability and energy storage performance of III-nitride compounds under electric field using first-principles methods. A nonpolar HE′ phase of (Al, Sc, B)N can be stabilized as the ground state at certain concentrations of the cations, and then transformed to MPE phase by application of an electric field. The evolution from PE to MPE yields a $P$-$E$ loop with a large polarization that saturates only at very large electric field and tiny polarization switching hysteresis. This peculiar $P$-$E$ response in nitrides is predicted to yield giant energy densities over 300 $J/cm^3$, greatly surpassing the record value of about 200 $J/cm^3$. The proposed design strategy and the specific material identified open the way towards dielectric capacitors with giant energy density and ultrahigh efficiency.

## Methods

### First-principles calculations

First principles calculations are performed via the Vienna ab initio simulation package (VASP)[30] within the framework of the projector augmented wave (PAW) potentials combined with PBEsol function. We use the plane wave energy cutoff of 550 eV[31,32]. The valence electron configuration is treated as following: Al $3s^23p^3$, B $2s^22p^1$, Sc $3s^23p^64s^23d^1$ and N $2s^22p^3$. The structures are fully relaxed until the force is converged to within 5 meV/Å on each atom. The spontaneous polarization is calculated by producing atomic displacement with respect to the paraelectric phases and Born effective charges[33]. The equilibrium structure under electric field is obtained by minimizing the electric enthalpy[34,35]. We use the scheme of electric enthalpy functional

$$F(R,\varepsilon) = E_{KS}^0(R) - P(R) \cdot \varepsilon, \qquad (4)$$

where $E_{KS}^0$ is the zero-field ground-state Kohn-sham energy at the coordinates $R$, $P$ is the polarization, and $\varepsilon$ is the electric field. Under an applied electric field, the equilibrium coordinates that minimize the electric enthalpy function generally satisfy the force-balance equation $-\frac{dE_{KS}^0}{dR} + Z^0 \cdot \varepsilon = 0$, where the $Z^0$ represents the zero-field Born effective change tensor. It has been repeatedly observed that, as compared with experiment, relatively large electric field values have to be used in first-principles calculations[4], hence, here the electric field values in the *P-E* loops were rescaled by a factor of 1/2 in this study to consist with measurements[19,21] (see Fig. S2).

**Structural configurations**

In order to mimic the cations random distribution of (Al, Sc, B)N, the structural configurations are firstly decided by the special quasirandom structure (SQS) method using the mcsqs code implemented in alloy theoretic automated toolkit (ATAT)[36,37]. Over twenty configurations are considered by SQS methods. We find different configurations may give large energy difference and therefore B cation distributions could greatly affect the stability. The lowest energy configuration by SQS calculations is selected, and then we construct structures with similar B-B connection (the bonding of B cations) with the lowest energy structure from SQS. (see section III in Supporting Information). The final *P-E* loop for one concentration of (Al, Sc, B)N are obtained by averaging the calculations of the low energy structures.

## Data availability

The source data for Figs. 3-5 in this study are provided in the Source Data file. Source data are provided with this paper.

## Code availability

All DFT calculations were performed with VASP, which is proprietary software for which the Yang's lab owns a license.

## Acknowledgements

The authors thank the National Key R&D Program of China (Grant Nos. 2020YFA0711504 and 2022YFB3807601), and the National Science Foundation of China (Grants No. 12274201, No. 52232001, No. 51721001, No. 52003117). J. Í.-G. acknowledges the financial support from the Luxembourg National Research Fund through Grant No. C21/MS/15799044/FERRODYNAMICS. We are grateful to the HPCC resources of Nanjing University for the calculations.

## Author contributions

Y.Y. and D.W. conceived and designed the research. Z.L. and X.M. performed the theoretical calculations supervised by YY. The results are analyzed by Z.L., X.M., L.C., X.Y., J-M.L., C-G.D., J.I.-G., D.W., and Y.Y. All authors followed the development of the research, discussed the results and contributed to the preparation of the manuscript.

## Competing interests

The authors declare no competing interests.

## Tables

Table I. The fitting coefficients of *P-E* loops by Eq. (2) and energy density (*W*) calculated by first principle and Eq. (3).

| Structure | Fitting coefficients | | | | $W$ (J/cm$^3$) | |
| --- | --- | --- | --- | --- | --- | --- |
| | $a$ (N·m$^2$/C$^2$) | $b$ ($10^{-6}$· N·m$^6$/C$^4$) | $c$ ($10^{-9}$· N·m$^{10}$/C$^6$) | $d$ ($10^{-13}$· N·m$^{14}$/C$^8$) | DFT | Eq. (3) |
| PE-MPE | 0.045 | 0.479 | -0.268 | 0.163 | 308 | 309 |
| PE-FE | 0.037 | -4.02 | 0.251 | -0.050 | 204 | 222 |
| RFE | 0.012 | 2.94 | -0.261 | 0.125 | 194 | 194 |

## Figure Captions

**Figure 1. Schematic diagram of *P-E* loops of dielectric materials**. **a** FE, **b** RFE, **c** AFE-FE, and **d** PE-MPE phase transition under electric field. The blue areas represent energy density $W$. $E_0$ represents phase transition field. In panel **d**, the polarization

increases linearly and slowly at low electric field; then, the system enters a second linear regime (MPE region) where the polarization grows more quickly, to eventually yield a large saturation polarization that occurs only at large fields.

**Figure 2. Structure for nitride phases**. **a** WZ, **b** HE, **c** BB, **d** RS, **e** ZB phase. Sliver spheres represent cation ions (Al, Sc, B), gray spheres represent N ions.

**Figure 3. Stability and phase diagram of (Al, Sc, B)N**. **a** The formation enthalpy $\Delta H$ of (Al, Sc, B)N system as a function of Sc concentration with B concentration of 22%. **b** Ternary phase diagram of (Al, Sc, B)N system. **c** Schematic of the total energy as a function of polarization for PE HE′ and ferroelectric FE WZ′ phases as the Sc concentration increases. **d** The atomic structures for WZ′, HE′, and RS′ phase, which similar to WZ, HE, and RS phase, respectively, with lattice distortions around B ions. The silver, magenta and green spheres represent Al, Sc and B atoms, respectively. The gray spheres represent N atoms. The point A presents the concentration of Al, Sc and B are 25%, 50% and 25%, respectively.

**Figure 4. The *P-E* loop and energy storage performance of (Al, Sc, B)N**. The *P-E* loops for (Al, Sc, B)N with Sc concentration of **a** 59% and **b** 50% for B concentration $y$ = 13%, 16%, 19%, 22%, 25%. The applied electric field maximum is 6 MV/cm. **c** The energy density $W$ and efficiency $\eta$ as a function of B concentration for Sc concentration from 47% to 59%. **d** The energy density of (Al, Sc, B)N in possible concentration of Sc and B. $P$ represents polarization change under electric filed comparing to the initial structure under zero field. Open and solid symbols indicate the polarization during the charging and discharging processes, respectively.

**Figure 5. PE-FE, RFE and PE-MPE *P-E* loops and the corresponding atomic structures. a** *P-E* loops for PE-FE (blue), RFE (green), and PE-MPE (red). Blue and red colors represent the *P-E* loop for $(Al_{0.41}Sc_{0.59})N$ and $(Al_{0.28}Sc_{0.59}B_{0.13})N$, respectively. Green color represent the experimental results of $Bi(Mg_{0.5}Ti_{0.5})O_3$-$SrTiO_3$-based RFE films from Ref. [8]. The open and solid symbols represent the charging and discharging behaviors, respectively, where the data is from our first-principles calculations or experiment results of Ref. [8]. The solid lines represent fittings using Eq. (2). Note some open symbols is overlaid by solid ones. **b** and **c** show, respectively, the atomic structures of the $(Al_{0.41}Sc_{0.59})N$ and $(Al_{0.28}Sc_{0.59}B_{0.13})N$ compounds under electric fields, corresponding to the data shown in panel **a**. The silver, magenta and green spheres represent Al, Sc and B atoms. The gray spheres represent N atoms.

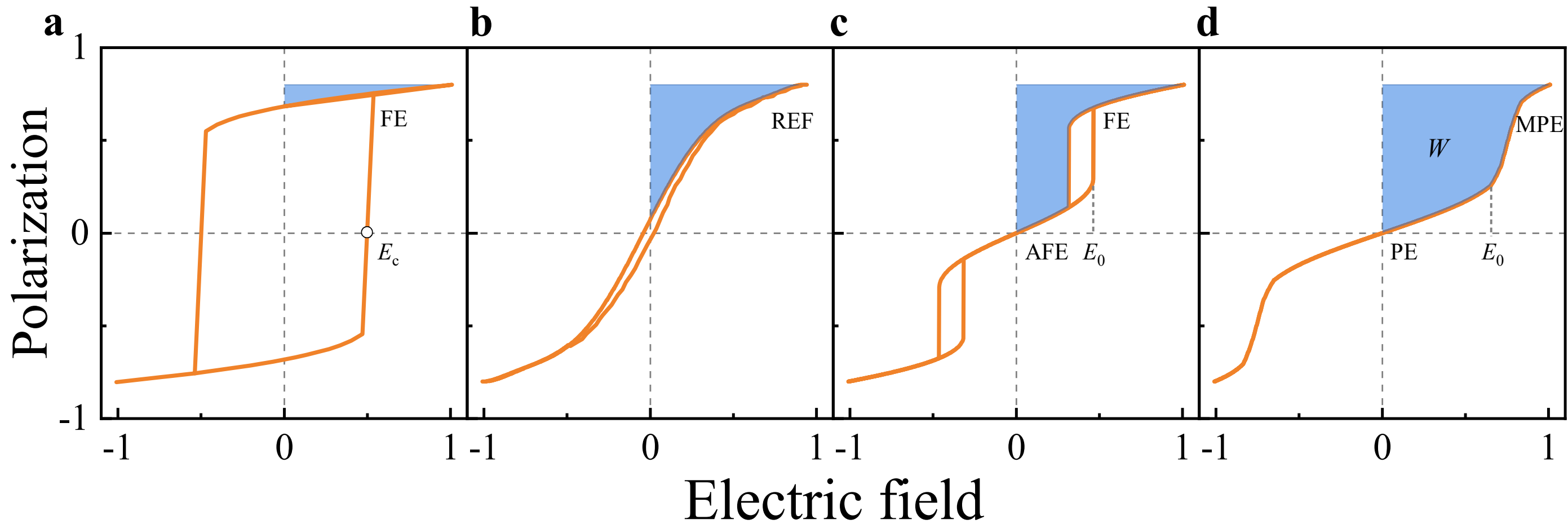

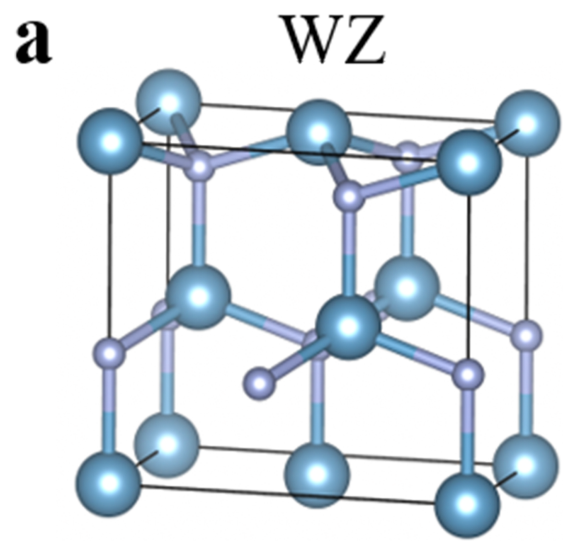 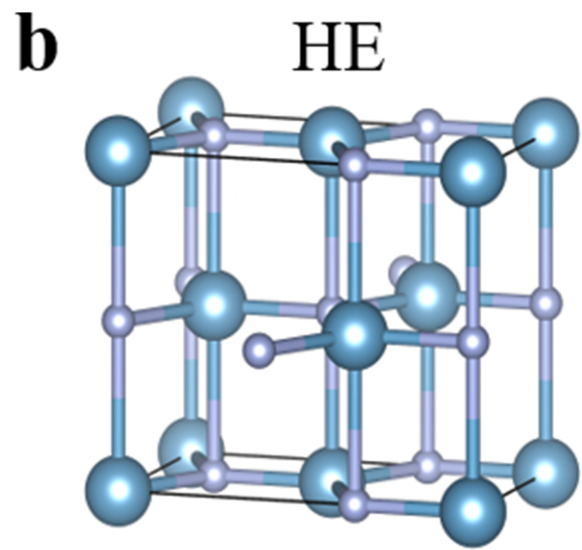 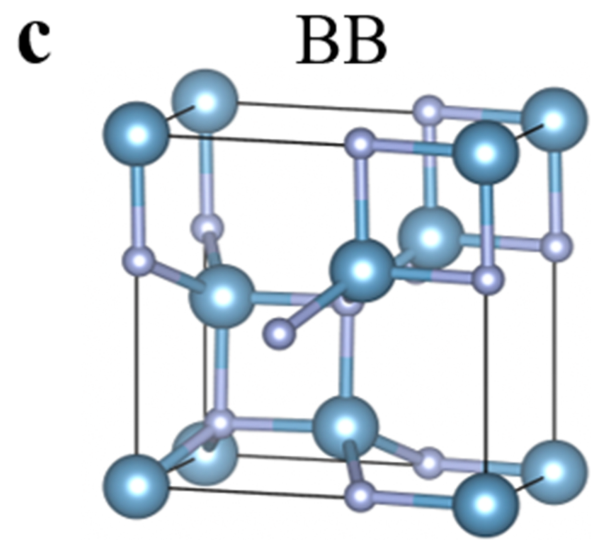 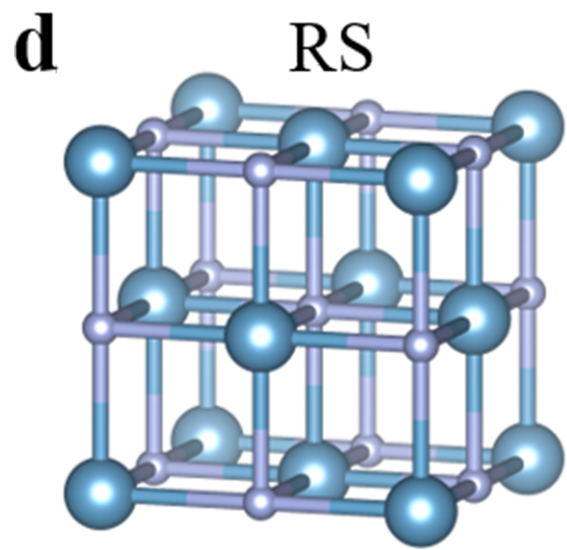 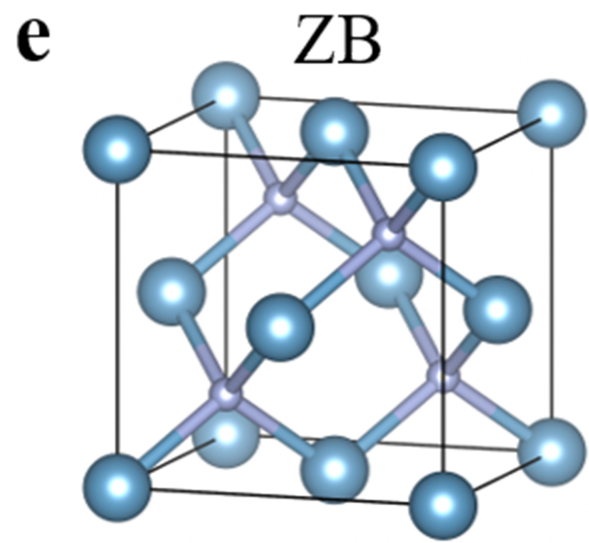 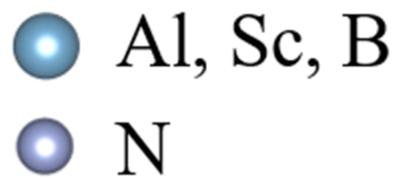 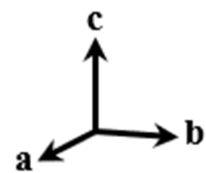

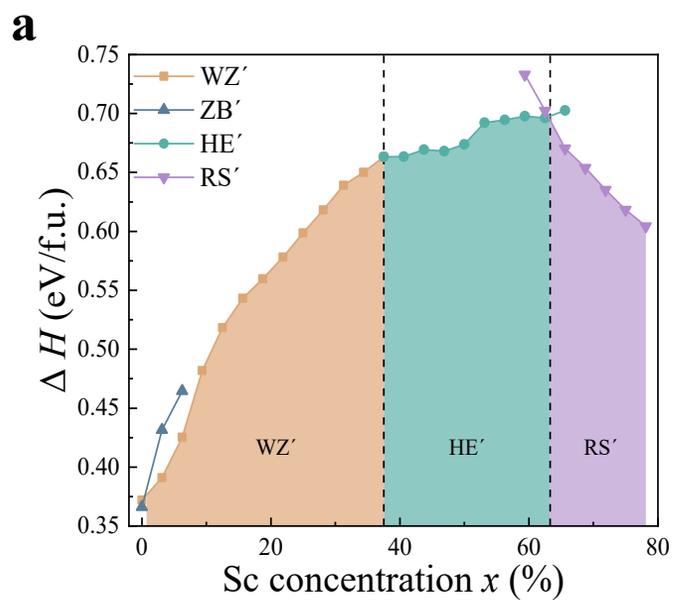
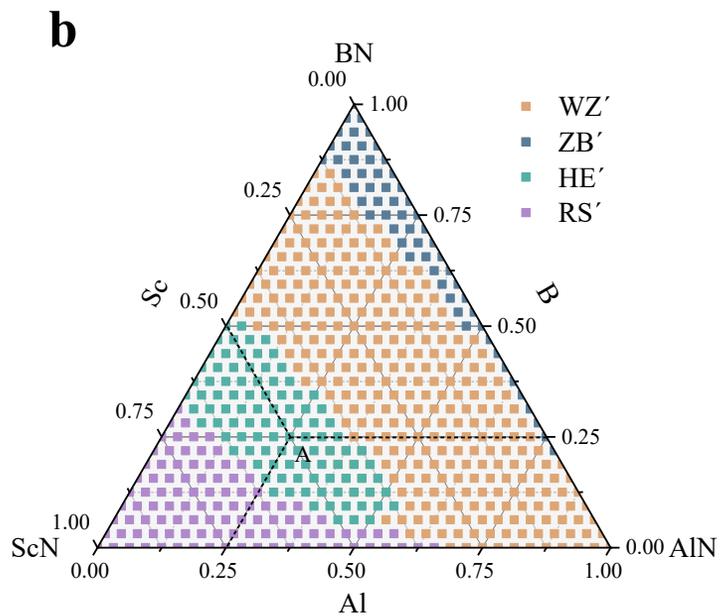
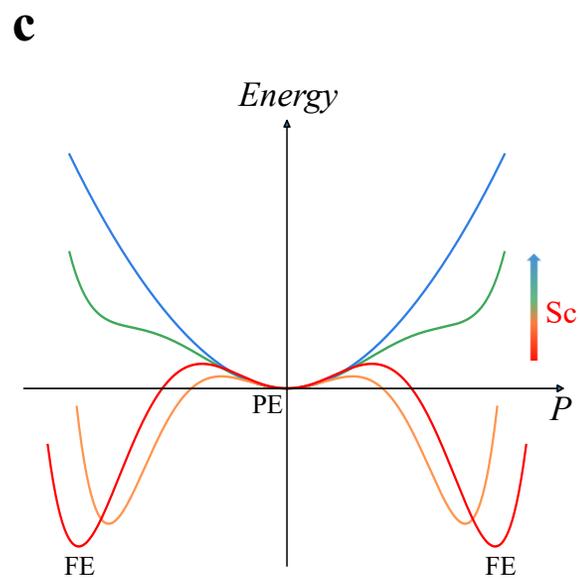
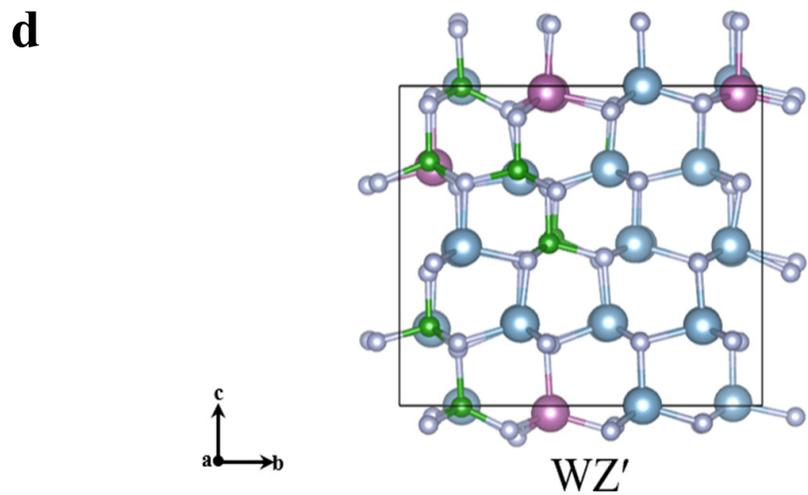

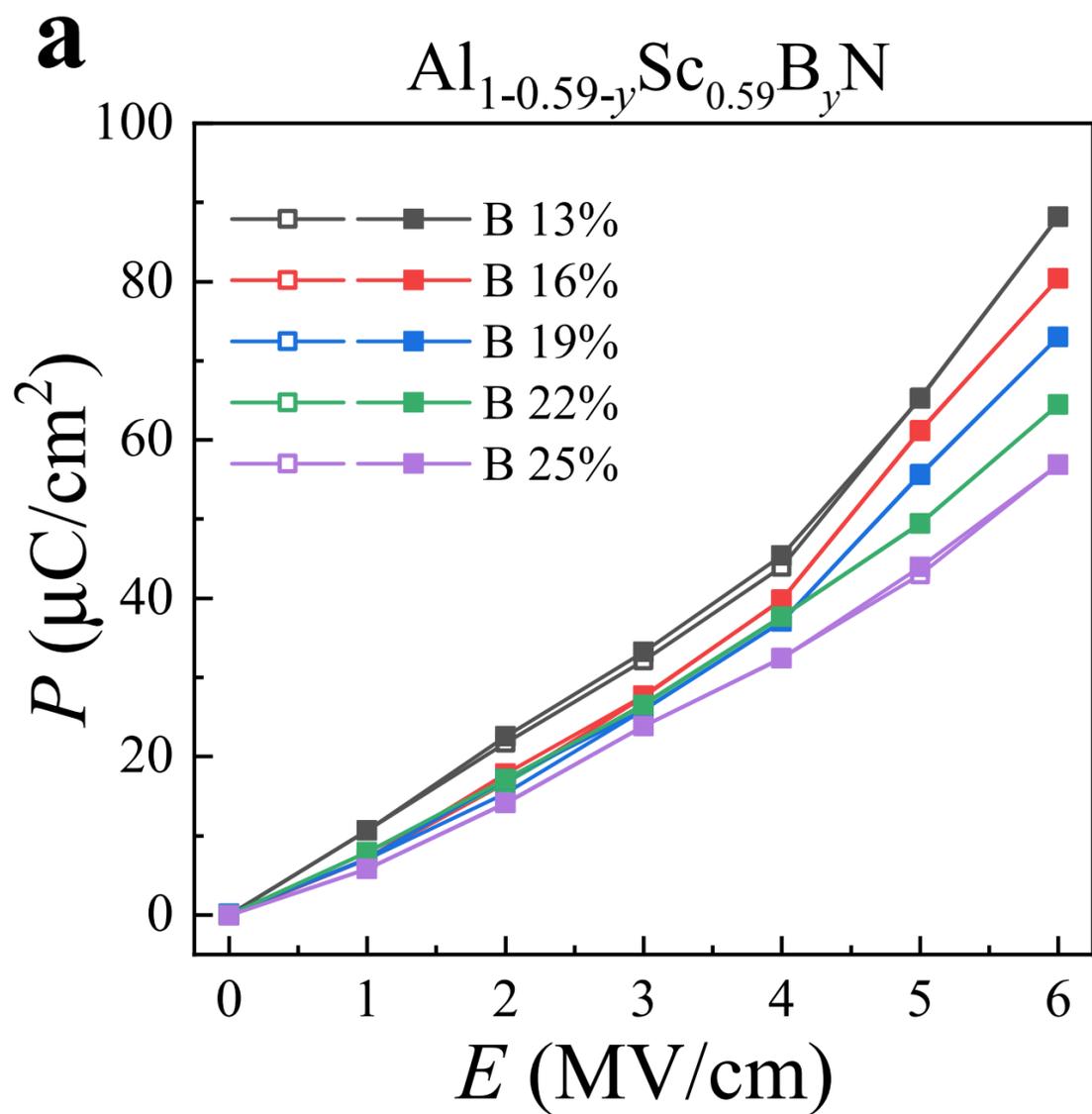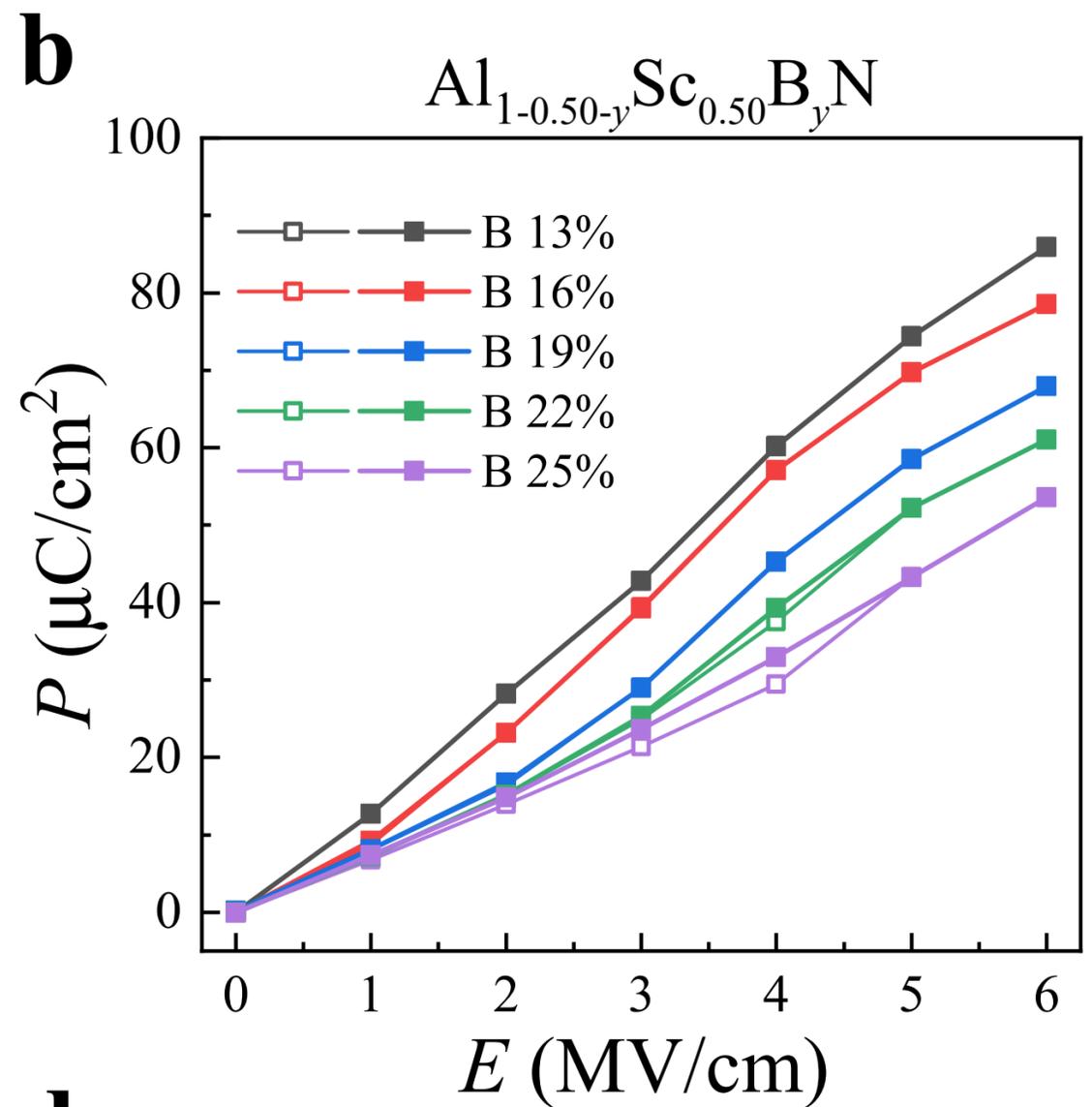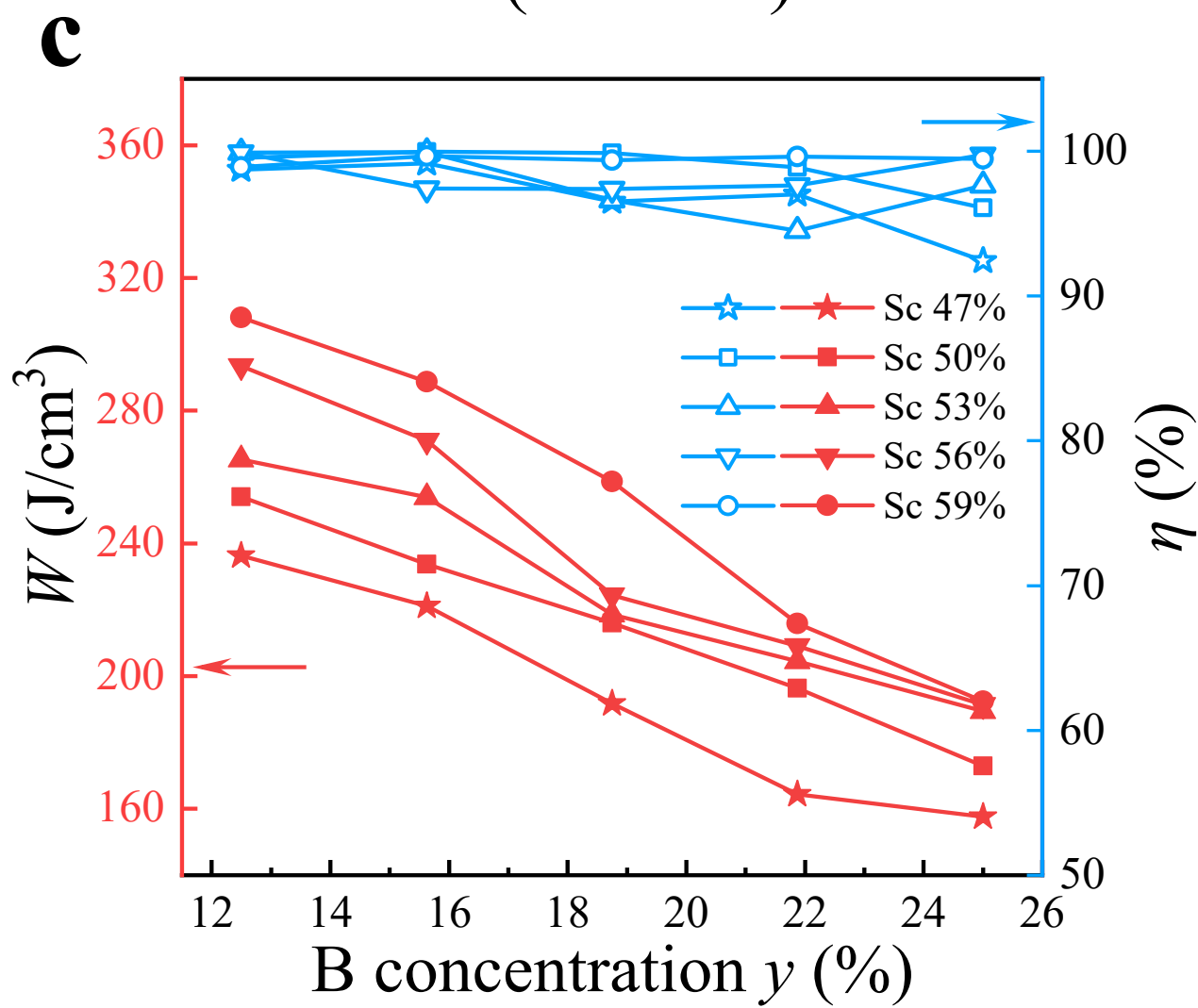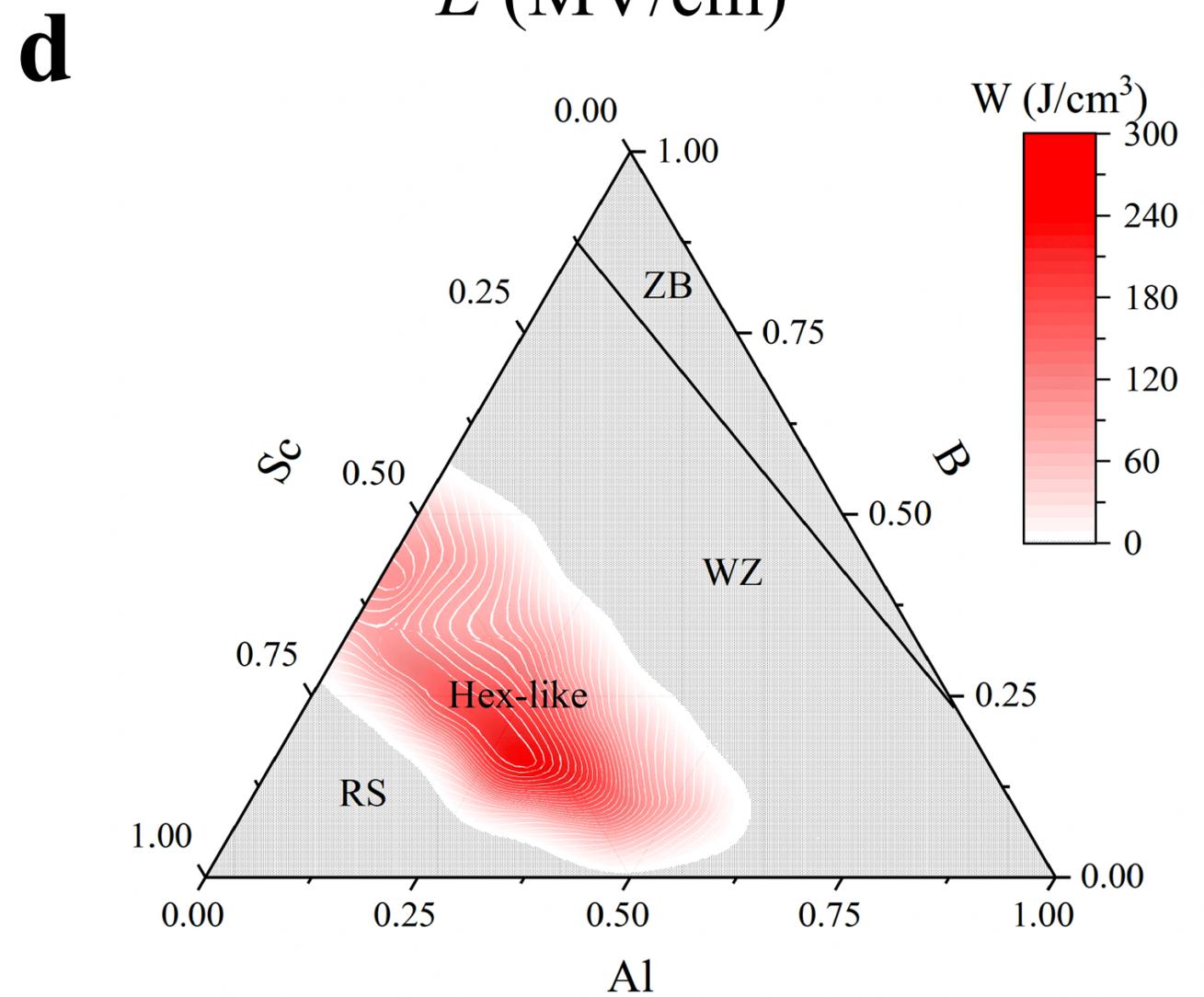

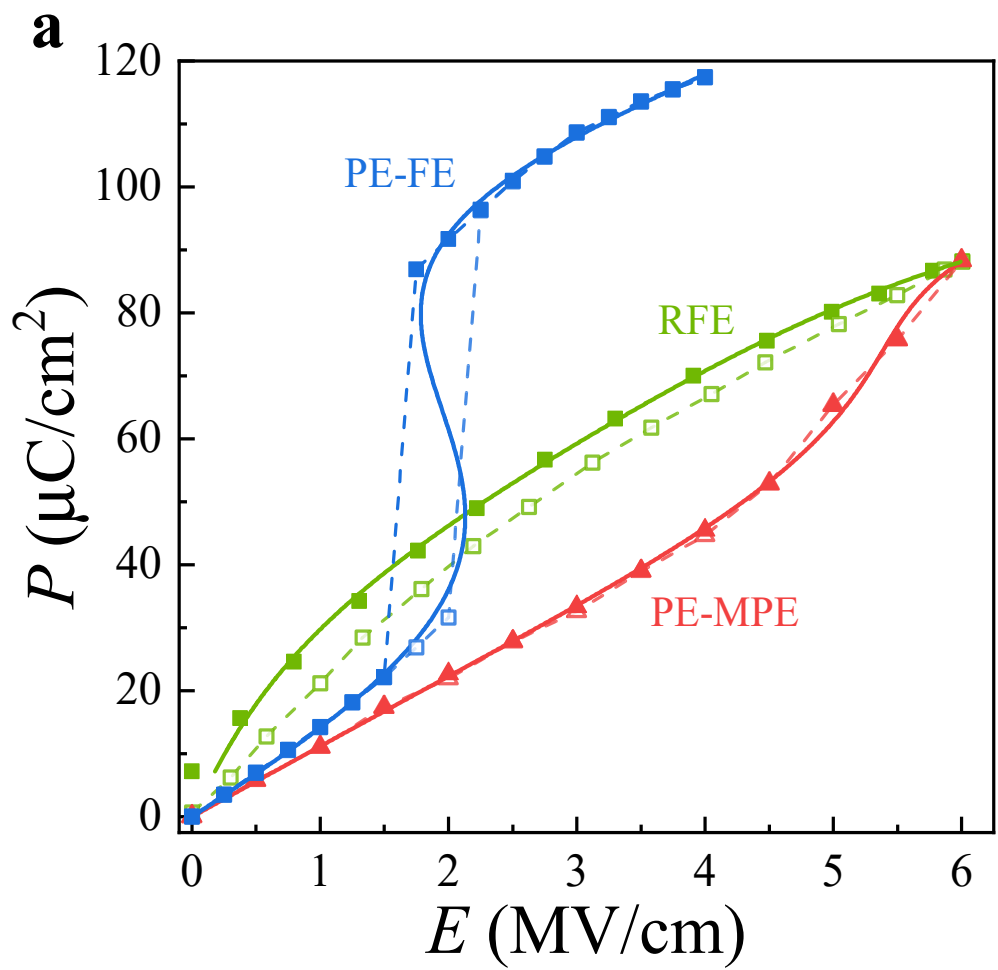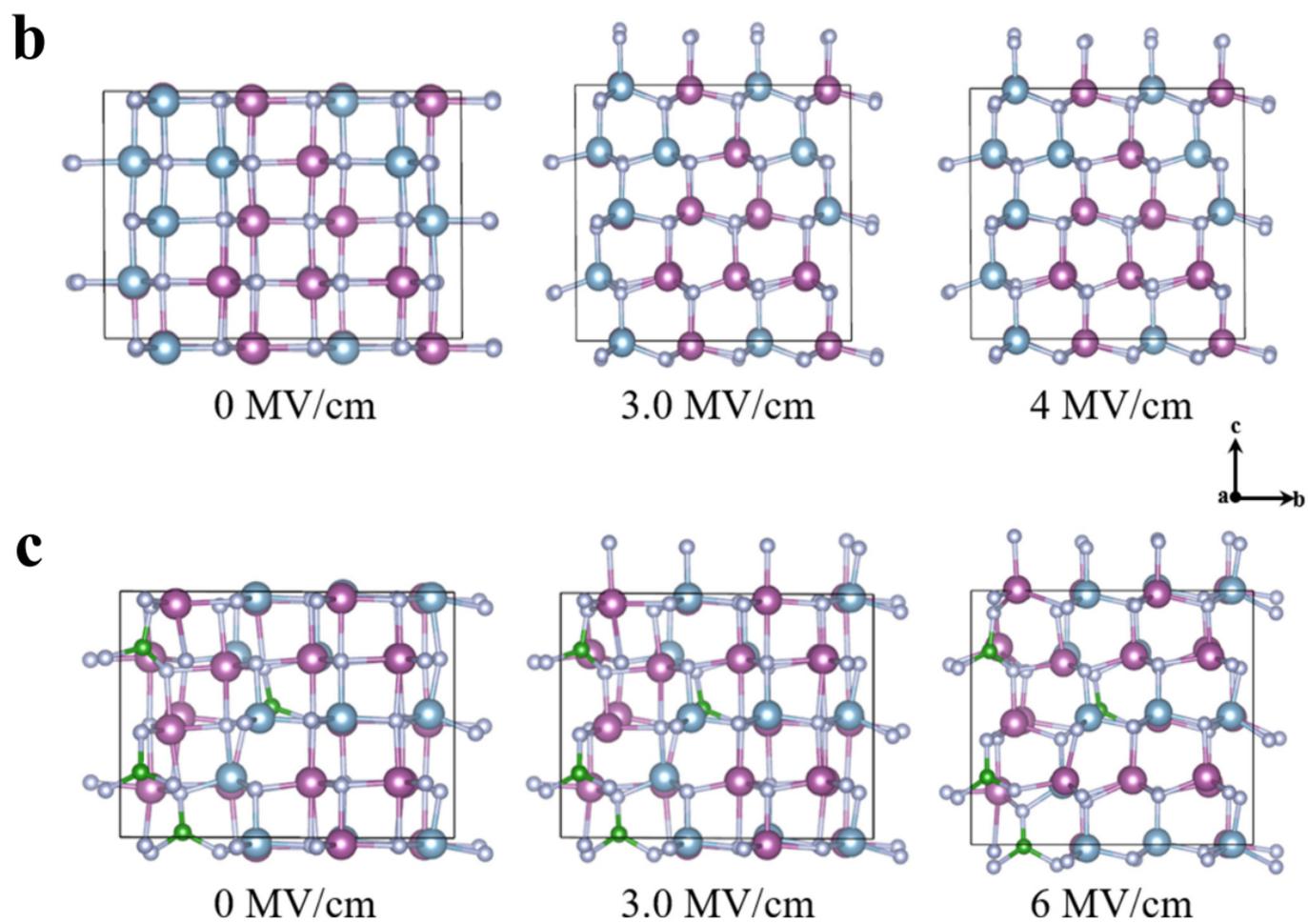